\documentclass[final,5p,times,twocolumn,sort&compress,10pt]{elsarticle}

\usepackage{amssymb}
\usepackage{lipsum}
\usepackage{mathptmx} 
\usepackage{mathrsfs}
\usepackage{amsthm}
\usepackage{amsmath}
\usepackage{amsfonts}
\usepackage{comment}
\usepackage{multirow,multicol}
\usepackage{subcaption}
\usepackage{graphicx,dblfloatfix}
\usepackage{subfig}
\usepackage{adjustbox}
\usepackage{tabularx}
\usepackage{multirow,multicol}
\usepackage{enumitem}
\usepackage{float}
\usepackage{stackengine} 
\usepackage{comment}
\usepackage{bm}
\usepackage{color}
\usepackage{url}
\usepackage[colorlinks]{hyperref}
\usepackage{cleveref}
\usepackage{parskip}
\usepackage{setspace}
\usepackage{placeins}
\usepackage{xkeyval}
\usepackage[normalem]{ulem}
\usepackage[margin=1cm]{caption}
\makeatletter
\providecommand{\sf@counterlist}{} 
\makeatother

\newcommand{\rbra}[1]{\langle #1||}
\newcommand{\rket}[1]{||#1\rangle}

\journal{Physics Letters B}

\begin{document}

\begin{frontmatter}



\title{A pathway to unveiling neutrinoless $\beta\beta$ decay nuclear matrix elements via $\gamma\gamma$ decay }

\author[1,2,3,4]{Beatriz Romeo}
\author[5,6]{Damiano Stramaccioni}
\author[2,3]{Javier Men\'{e}ndez} 
\author[5]{Jose Javier Valiente-Dob\'{o}n}

\affiliation[1]{organization={Department of Physics and Astronomy, University of North Carolina},
            postcode={27516-3255}, 
            city={Chapel Hill},
            country={USA}}

\affiliation[2]{organization={Departament de Física Quàntica i Astrofísica, Universitat de Barcelona},
            postcode={08028}, 
            city={Barcelona},
            country={Spain}}

\affiliation[3]{organization={Institut de Ciències del Cosmos, Universitat de Barcelona},
            postcode={08028}, 
            city={Barcelona},
            country={Spain}}            

\affiliation[4]{organization={Donostia International Physics Center},
            postcode={20018}, 
            city={San Sebastián},
            country={Spain}}        

\affiliation[5]{organization={Laboratori Nazionali di Legnaro, Istituto Nazionale di Fisica Nucleare},
            postcode={35020}, 
            city={Legnaro},
            country={Italy}}

\affiliation[6]{organization={Dipartimento di Fisica e Astronomia, Università degli Studi di Padova},
            postcode={35121}, 
            city={Padova},
            country={Italy}}

\begin{abstract}
We investigate the experimental feasibility of detecting second-order double-magnetic dipole ($\gamma\gamma$-$M1M1$) decays from double isobaric analog states (DIAS), which have recently been found to be strongly correlated with the nuclear matrix elements of neutrinoless $\beta\beta$ decay. Using the nuclear shell model, we compute theoretical branching ratios for $\gamma\gamma$-$M1M1$ decays and compare them with other competing processes, such as single-$\gamma$ decay and proton emission, which represent the dominant decay channels. We also estimate the potential competition from internal conversion and internal pair creation, which can influence the decay dynamics. Additionally, we propose an experimental strategy based on using LaBr$_3$ scintillators to identify $\gamma\gamma$-$M1M1$ transitions from the DIAS amidst the background of the competing processes. Our approach emphasizes the challenges of isolating the rare $\gamma\gamma$-$M1M1$ decay and suggests ways to enhance the experimental detection sensitivity. Our simulations suggest that it may be possible to access experimentally $\gamma\gamma$-$M1M1$ decays from DIAS, shedding light on the neutrinoless $\beta\beta$ decay nuclear matrix elements.

\end{abstract}







\end{frontmatter}




\section{Introduction}
\label{introduction}
The neutrinoless $\beta\beta$ ($0\nu\beta\beta$) decay of a nucleus---transmuting two neutrons into protons and emitting only two electrons---while currently hypothetical, would have profound consequences in fundamental physics~\cite{Agostini:2022zub,Gomez-Cadenas:2023vca}. First, it would establish that neutrinos are their own antiparticle. 
In addition, it would be the first observation in the laboratory of a process breaking the balance of matter and antimatter, which may help to understand the dominance of matter observed in the universe. Further, it can give hints on the value of the absolute neutrino masses.
A portfolio of experiments pursue the detection of $0\nu\beta\beta$ decay using various nuclei, mainly using $^{76}$Ge, $^{82}$Se, $^{100}$Mo, $^{130}$Te and $^{136}$Xe~ \cite{Anton2019,Agostini2020,Majorana:2022udl,Azzolini2022,CUPID-Mo:2023,Adams2024,Abe2024,Agrawal:2024zfv}, but other isotopes such as $^{48}$Ca are considered as well~\cite{CANDLES:2020iya}.    

The $0\nu\beta\beta$ rate depends, among other terms, on nuclear matrix elements that need to be calculated, as the decay has not been observed yet~\cite{Engel:2016xgb}. Even the best current calculations disagree by factors of a few~\cite{Yao:2019rck,Wirth:2021pij,Novario:2020dmr,Belley2023heavy,Belley2024,Weiss:2021rig,Coraggio:2020hwx,Coraggio:2022vgy,Jokiniemi:2022ayc,Horoi:2022ley,Horoi:2023uah,Lv:2023dcy,Simkovic:2018hiq,Mustonen:2013zu,Fang:2018tui,Rodriguez:2010mn,LopezVaquero:2013yji,Ding:2023dnl,Yao:2021wst}, and recent attempts to quantify the theoretical uncertainties also suggest large errors~\cite{Gomez-Cadenas:2023vca}. Since the nuclear matrix elements are key not only to interpret a $0\nu\beta\beta$ signal, but also to plan future searches, multiple efforts have been proposed with the goal of better constraining their values.

One possible avenue is to use nuclear structure observables  related with $0\nu\beta\beta$ decay. These include, for instance, pair transfer amplitudes~\cite{Brown:2014yda}, ordinary muon capture~\cite{Jokiniemi:2019nne,Jokiniemi:2020ydy,Jokiniemi:2021qqg,Jokiniemi:2024zdl}, heavy ion charge-exchange reactions~\cite{NUMEN:2022ton}, double Gamow-Teller transitions~\cite{Shimizu:2017qcy,Rodriguez:2012rv,Wang:2024zkl,Jokiniemi_JM_2023,Santopinto:2018nyt,Yao:2022usd},  $\beta\beta$ decay with the emission of two neutrinos~\cite{Jokiniemi:2022ayc,Horoi:2022ley,Horoi:2023uah} or proton-neutron phase-shifts at moderate energies~\cite{Belley:2024zvt}.   
Other processes related to $0\nu\beta\beta$ decay are electromagnetic transitions. Already several decades ago, $\gamma$ decays from isobaric analog states (IAS) were proposed to gain insight on electroweak transitions~\cite{Ejiri68}: in particular the electric dipole decay from the IAS of $^{141}$Pr informed the first-forbidden $\beta$ decay of $^{141}$Ce. Very recently, a similar idea was proposed to relate $0\nu\beta\beta$ and $\gamma$ decays~\cite{Ejiri_2023}. More broadly, many works have exploited the common spin-isospin nature of the strong, weak and electromagnetic forces in order to provide detailed analyses of nuclear structure~\cite{Fujita2011}, nuclear responses~\cite{Fujita2011} and neutrino-nuclear responses~\cite{Ejiri2019}. 

A particularly interesting correlation between the electromagnetic sector and $0\nu\beta\beta$ decay was proposed in Ref.~\cite{Romeo_2022}. 
With focus on second-order decays---just like $0\nu\beta\beta$---this work found that $\gamma\gamma$ double-magnetic dipole ($M1M1$) nuclear matrix elements from the transition of double isobaric analogue states (DIAS) to the ground state (GS) are very well correlated with $0\nu\beta\beta$ nuclear matrix elements, as long as the two emitted photons have similar energy. The correlation between $\gamma\gamma$-$M1M1$ and $0\nu\beta\beta$ nuclear matrix elements, initially obtained for the nuclear shell model, was also recently observed with a very different many-body approach, the quasiparticle random-phase approximation (QRPA) method~\cite{Jokiniemi_JM_2023}. 
This correlation between $\gamma\gamma$ and $0\nu\beta\beta$ nuclear matrix elements presents some advantages. For instance, since both are second-order processes, there is no need to combine two first-order ones guessing unknown phases, like in muon capture or single-$\gamma$ decays~\cite{Jokiniemi:2020ydy,Ejiri_2023}. Moreover, electromagnetic reactions are easier to model than those mediated by the strong force~\cite{Lenske:2019iwu} relevant for double Gamow-Teller transitions. Finally, in the electromagnetic sector calculations may not face the overestimation---``quenching''---that many approaches observe for two-neutrino $\beta\beta$ nuclear matrix elements~\cite{Agostini:2022zub,Gysbers:2019uyb}. 

From the nuclear structure point of view, exploring DIAS and isovector electromagnetic transitions is also interesting, as it opens new paths to investigate isospin phenomena and rare decays in nuclei. Indeed, even though $\gamma\gamma$ transitions were first measured decades ago~\cite{Kramp:1987dm}---only in cases where single-$\gamma$ emission is forbidden---the last decade has seen a revival of $\gamma\gamma$ measurements~\cite{Walz15,Söderström2020,Freire-Fernandez:2023ajk}, including competition with $\gamma$ decay. 
Nonetheless, $\gamma\gamma$ transitions from DIAS have not been measured yet, and possible experiments in medium-mass nuclei such as those used in $0\nu\beta\beta$ searches present mainly two challenges. Firstly, to populate efficiently the unbound DIAS. Secondly, to measure the very suppressed $\gamma\gamma$ branch, since the $\gamma$ decay and particle emission channels are open. 

\begin{figure}[t]
    \centering
    \includegraphics[width=
    0.48\textwidth]{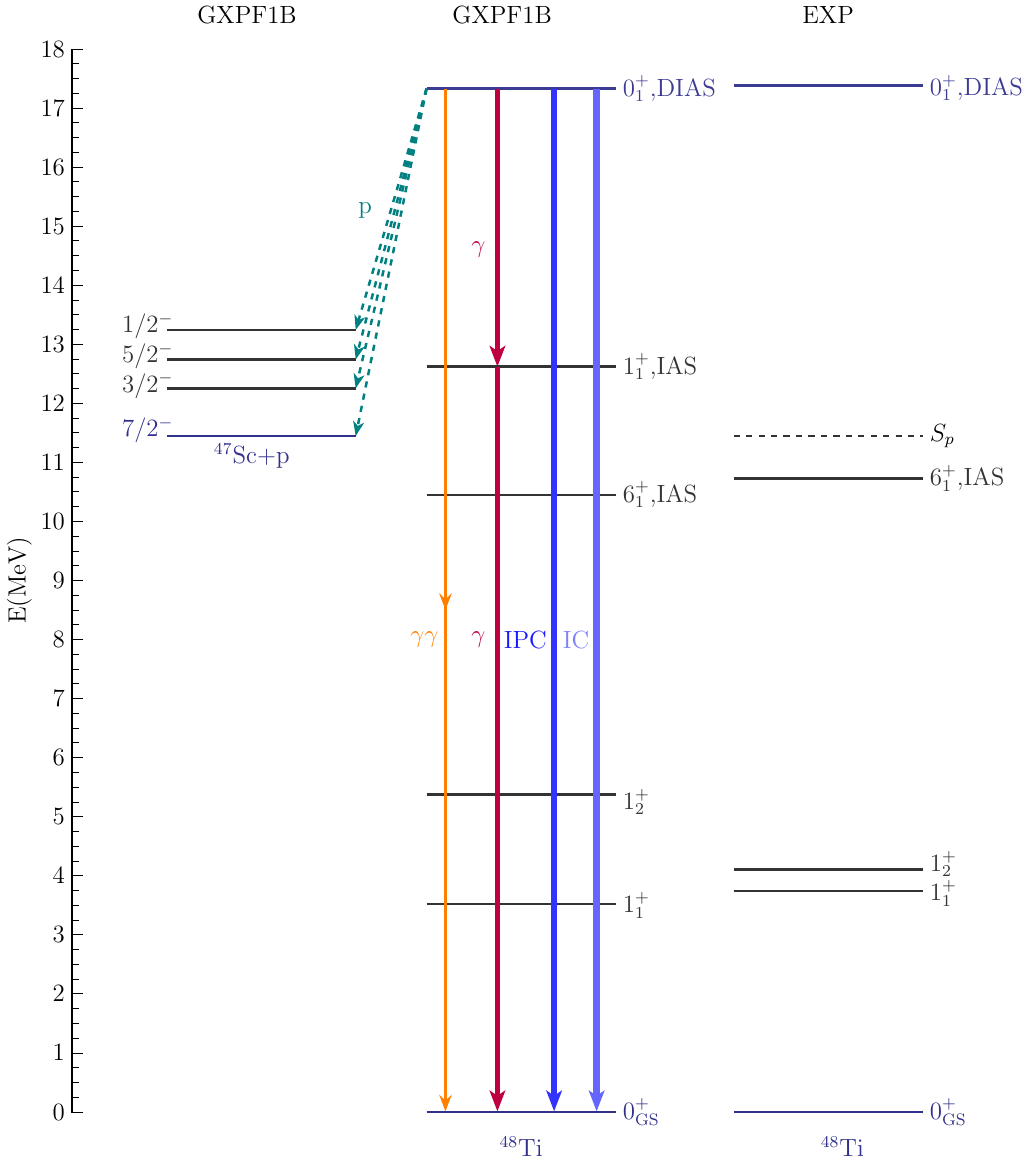}
\caption{Decay scheme of the DIAS of $^{48}$Ti, including proton (p) emission, single-$\gamma$ and $\gamma\gamma$ decays, internal conversion (IC) and internal pair creation (IPC). The shell-model energies of selected $^{48}$Ti low-energy states, obtained using the GXPF1B interaction, are compared to experiment~\cite{NucData22}.}
    \label{fig:levScheme}
\end{figure}

In this manuscript we study the second aspect with a complementary theoretical and experimental analysis. We focus on the $0\nu\beta\beta$ daughter nuclei expected in $0\nu\beta\beta$-decay experimental searches: $^{48}$Ti, $^{76}$Se, $^{82}$Kr,  $^{130}$Xe and  $^{136}$Ba. From the theory side, we fully characterize the decay width and branching ratios of $\gamma\gamma$-$M1M1$ transitions from DIAS based on nuclear shell-model calculations and systematics. From an experimental point of view, based on the theory predictions, we present a strategy to identify the DIAS $\gamma\gamma$-$M1M1$ decays from the competing decay channels using current detector technologies. As competing processes, we consider single-$\gamma$ decay, proton emission, internal pair creation (IPC) and internal conversion (IC). Figure~\ref{fig:levScheme} shows the corresponding schematic decay scheme for the  DIAS in $^{48}$Ti. 

\section{Double-magnetic dipole transitions from DIAS}
We study the second-order $\gamma\gamma$ nuclear transitions, 
\begin{align}
^{A}_{Z}Y^{\text{DIAS}}_{N}\longrightarrow\, ^{A}_{Z}\!Y_{N}+2\gamma\,,
\end{align}
where the two photons, emitted simultaneously, share the energy difference from the initial DIAS, $0^+_\text{DIAS}$, to the final GS, $0^+_\text{GS}$. Here we focus on nuclei which are the final states of measured $\beta\beta$ decays, all of which have total angular momentum and parity $J^P=0^+$. In the transition, the isospin changes in two units: from $T_\text{DIAS}=T_z+2$ to $T_\text{GS}=T_z$, with $T_z=(N-Z)/2$ given by the difference between the nucleus neutron and proton numbers, and $A=N+Z$.


Because of angular momentum and parity conservation in electromagnetic transitions, the leading $0^+_\text{DIAS}\rightarrow 0^+_{GS}$ $\gamma\gamma$ contribution comes from double-electric dipole ($E1E1$) and double-magnetic dipole ($M1M1$) decays. However, only the latter are of interest from the point of view of the $\beta\beta$-$\gamma\gamma$ correlation~\cite{Romeo_2022}, because only the magnetic dipole operator,
\begin{equation}
    \mathbf{M1}=\mu_N\sqrt{\frac{3}{4\pi}}\sum_{i=1}^A(g_i^l\mathbf{l}_i+g_i^s\mathbf{s}_i),\label{eq:magnetic_dipole_def}
\end{equation}
expressed in terms of the angular-momentum ($\mathbf{l}$) and spin ($\mathbf{s}=\frac{1}{2}\mathbf{\sigma}$) operators, has a structure similar to the Gamow-Teller one that drives $0\nu\beta\beta$ decay. We use bare neutron ($n$) and proton ($p$) orbital and spin $g$-factors $g_l^{(n)}=0$, $g_l^{(p)}=1$, $g_s^{(n)}=-3.826$, $g_s^{(p)}=5.586$, and $\mu_N$ is the nuclear magneton.

The $\gamma\gamma$-$M1M1$ transition amplitude is proportional to the generalized nuclear polarizability~\cite{Kramp}:
\begin{align}
   P_0(M1M1;E_1,E_2)=&\frac{4\pi\, E_1E_2}{3\sqrt{3}}\left[\alpha_{M1M1}(E_1)+\alpha_{M1M1}(E_2)\right],\label{eq:generalized_pol_0}\\
    \alpha_{M1M1}(E)=&\sum_n\frac{(0_{\text{GS}}^+||\mathbf{M1}||1_n^+)(1_n^+||\mathbf{M1}||0_{\text{DIAS}}^+)}{E_n-E_{\text{DIAS}}+E},\label{eq:alpha_M1M1}
\end{align}
where the reduced matrix elements involve the set of $1^+_n$ intermediate IAS, with energy $E_n$ and isospin $T_n=T_\text{GS}+1$, and $E_1$, $E_2$, $E_{\text{DIAS}}$ are the energies of the photons and the DIAS, respectively. 
For the most probable case where the two photons are emitted with the same energy, $E_1=E_2=Q_{\gamma\gamma}/2$, the polarizability depends on a single nuclear matrix element:
\begin{equation}
M^{\gamma\gamma}(M1M1)=\sum_{n}\frac{\rbra{0^+_\text{GS}}\mathbf{M1}\rket{1^+_n}\rbra{1^+_n}\mathbf{M1}\rket{0^+_{\text{DIAS}}}}{E_n-(E_\text{GS}+E_\text{DIAS})/2}\,,
\label{eq:ggM1M1_NME}
\end{equation}
which has been found to be very well correlated with the $0\nu\beta\beta$-decay one~\cite{Romeo_2022,Jokiniemi_JM_2023}. In practice, if the two photons have relatively similar energies, $\Delta\varepsilon=|E_1-E_2|\ll Q_{\gamma\gamma}$, the nuclear matrix element varies smoothly and the $\gamma\gamma$-$M1M1$ decay width is 
\begin{align}
    \Gamma_{\gamma\gamma}(&M1M1;\Delta\varepsilon)= \frac{8\pi}{243}M^{\gamma\gamma}(M1M1)^2 \times \nonumber\\ & \frac{\Delta\varepsilon}{Q_{\gamma\gamma}}\left[\frac{1}{2} Q_{\gamma\gamma}^7 -\frac{1}{2}\Delta\varepsilon^2Q_{\gamma\gamma}^5
    +\frac{3}{10}\Delta\varepsilon^4Q_{\gamma\gamma}^3-\frac{1}{14}\Delta\varepsilon^6Q_{\gamma\gamma}\right].\label{eq:ggM1M1_decay_width}
\end{align}
This expression demands an experimental setup optimized to detect photons with energies $E_\gamma\simeq E_{\rm{DIAS}}/2$, and therefore a prior measurement of $E_{\rm DIAS}$ is necessary. 

We calculate the $\gamma\gamma$ decays within the nuclear shell model, using the ANTOINE~\cite{FNowacki,MPinedo} and KSHELL~\cite{Shimizu_2013} codes with isospin-symmetric interactions in different configuration spaces: KB3G~\cite{PovesKB3G} and GXPF1B~\cite{HonmaGXPF} for $^{48}$Ti with a $^{40}$Ca core and the $0f_{7/2}$, $1p_{3/2}$, $0f_{5/2}$ and $1p_{1/2}$ single-particle orbitals;  GCN2850~\cite{Caurier08}, JUN45~\cite{HonmaJUN45} and JJ4BB~\cite{BrownJJ4BB} for $^{76}$Se and $^{82}$Kr with a $^{56}$Ni core and the $1p_{3/2}$, $0f_{5/2}$, $1p_{1/2}$ and $0g_{9/2}$ orbitals; and GCN5082~\cite{Caurier08} 
for $^{130}$Xe and $^{136}$Ba with a $^{100}$Sn core in the space comprising the orbitals $0g_{7/2}$, $1d_{5/2}$, $1d_{5/2}$, $2s_{1/2}$ and $0h_{11/2}$.  Orbitals are given in spectroscopic notation $nl_j$, where $n$, $l$, $j$ are the principal, orbital and total angular momentum quantum numbers. We use these shell-model configuration spaces and interactions for all calculations throughout this work.

Table~\ref{tab:M1M1_gamma-gamma_widths} presents theoretical $\Gamma_{\gamma\gamma}(M1M1)$ values obtained demanding that $\alpha_{M1M1}(E)$ differs from $M^{\gamma\gamma}(M1M1)$ in less than 5\%, which translates into photon-energy differences, $\Delta\varepsilon\approx(1-2)$~MeV. Therefore, such $\gamma\gamma$-decay measurements would probe the nuclear matrix element correlated with the $0\nu\beta\beta$-decay one. The widths are very narrow, $\Gamma_{\gamma\gamma}\sim(10^{-6}-10^{-7})$~eV, and typically differ by a factor of $5-10$ between different shell-model interactions. These predictions are about an order of magnitude below the ones obtained with the expression for $\gamma\gamma$ widths given in Ref.~\cite{Freire-Fernandez:2023ajk}, which considers the full range of $\gamma$ energies and assumes constant generalized polarizabilities squared---summed for both $M1M1$ and $E1E1$---across different nuclei and nuclear states. 
While most $E_\text{DIAS}=Q_{\gamma\gamma}$ are not known, for $^{48}$Ti the calculated energy is very close to the experimental value, $E_\mathrm{DIAS}^\text{exp}$=17.4~MeV~\cite{Kouzes}, especially for the GXPF1B interaction.


\begin{table}[t]
\begin{center}
\def\arraystretch{1.3}
\begin{tabular}{c c c c} 
\hline\hline
   Nucleus  & $H_\mathrm{{eff}}$ & $E_\mathrm{DIAS}$ (MeV) & $\Gamma_{\gamma\gamma} (\times 10^{-7}~ \text{eV})$ \\ 
\hline
$^{48}$Ti & KB3G & 16.1  & 2.0 \\
\hline
$^{48}$Ti & GXPF1B &  17.3 & 0.55 \\
\hline
$^{76}$Se & GCN2850 & 19.2 & 8.7 \\
\hline
$^{76}$Se & JUN45 & 21.2 & 15  \\
\hline
$^{76}$Se & JJ44B & 23.6 & 32 \\
\hline
$^{82}$Kr & GCN2850 & 21.6  & 12  \\
\hline
$^{82}$Kr & JUN45 & 22.9  & 12  \\
\hline
$^{82}$Kr & JJ44B & 26.6 & 27 \\
\hline
$^{130}$Xe & GCN5082 & 27.7 & 2.2 \\
\hline
$^{136}$Ba & GCN5082 & 29.1 & 17  \\
\hline  
\end{tabular}
\end{center}
\caption{Calculated $\gamma\gamma$-$M1M1$ decay widths ($\Gamma_{\gamma\gamma}$) and energies ($E_\text{DIAS}$) of the $0^+_\text{DIAS}$ for all nuclei studied in this work. We use the nuclear shell model with different effective interactions ($H_{\rm{eff}}$).}
\label{tab:M1M1_gamma-gamma_widths}
\end{table}


\section{Competing decay channels}

Given the narrow widths expected from Table~\ref{tab:M1M1_gamma-gamma_widths}, the measurement of $\gamma\gamma$-$M1M1$ transitions demands a careful study of the competing processes: IC, IPC, single-$\gamma$ decay and nucleon emission.

\subsection{Internal conversion and internal pair creation}
\label{sec:IC}
In nuclei with $0^+$ first-excited states, IC and IPC are the main competing channels with $\gamma\gamma$ decay~\cite{Kramp:1987dm}. In general, these two processes are also expected to be relevant decay branches of the DIAS. The corresponding widths can be factorized into electron and nuclear parts~\cite{PhysRev.103.1035}:
\begin{equation}\Gamma_\text{IC/IPC} = \Omega_\text{IC/IPC}(E0, Q) \frac{|\langle \bar{r}^2 \rangle|^2}{R^4},
\end{equation}
where the ratio involving the mean squared and average nuclear radii, $\bar{r}^2$ and $R$, is order $\sim 1$. Thus, Table~\ref{tab:IC_IPC} presents the IC and IPC decay widths for $Q_\text{E0}=E_\text{DIAS}$ given by the $\Omega$ expressions derived by Church and Wesener~\cite{PhysRev.103.1035} and Wilkinson~\cite{WILKINSON19691}, respectively. The decays are relatively wide, especially for IPC in heavy nuclei. However, the estimates in Table~\ref{tab:IC_IPC} do not consider the finite-size of the nucleus or atomic screening effects, and for IC they only take into account the dominant K-shell. 

\begin{table}[b]
\begin{center}
\def\arraystretch{1.3}
\begin{tabular}{c c c c} 
\hline\hline
   Nucleus  & $Q_\text{E0}$ ($\mathrm{MeV}$) &  $\Gamma_{\mathrm{IPC}}(\mathrm{eV})$ & $\Gamma_{\mathrm{IC}}(\mathrm{eV})$ \\ 
\hline
  $^{48}$Ti  & 17.4 & 0.05  &  2$\cdot$10$^{-6}$  \\
\hline
  $^{76}$Se  &  21.3 & 0.3  &  3$\cdot$10$^{-5}$   \\
\hline
  $^{82}$Kr  &  23.7 & 0.5  &  6$\cdot$10$^{-5}$ \\
\hline
  $^{130}$Xe  &  27.7 & 2  &  0.001   \\
\hline
  $^{136}$Ba  & 29.1  & 3  &  0.002   \\  
    \hline\hline
\end{tabular}
\end{center}
\caption[]{Internal pair creation ($\Gamma_{\mathrm{IPC}}$) and internal conversion ($\Gamma_{\mathrm{IC}}$) decay widths for the $0^+_\text{DIAS}\rightarrow 0^+_{GS}$ transition, obtained from systematics~\cite{PhysRev.103.1035,WILKINSON19691} with $Q_\text{E0}=E_\text{DIAS}$.}
\label{tab:IC_IPC}
\end{table}

\subsection{First-order electromagnetic transitions}
Given the high energy of the DIAS, this state can decay via a first-order $\gamma$ transition. Dipole decays are expected to be dominant, but in our shell-model configuration spaces we cannot access a reliable set of $1^-$ states because this would demand including at least to harmonic oscillator shells~\cite{Togashi:2018dau}. We thus focus on magnetic dipole ($M1$) transitions, which lead to a decay width
\begin{equation}
    \Gamma_{\gamma}=\frac{4E^3_{\gamma}}{3}|(1_n^+||\mathbf{M1}||0_{\rm{DIAS}}^+)|^2.
\label{eq:single_gamma_width}
\end{equation}

Table~\ref{tab:M1_single-gamma_widths} lists the calculated decay widths obtained with the nuclear shell model, together with the energies for the dominant $1^+$ states and their branching ratio with respect to all $M1$ decays, also including those not listed in Table~\ref{tab:M1_single-gamma_widths}. The total decays widths are large and have a relatively similar value for all nuclei. However, while for $^{48}$Ti the theoretical calculations predict that a single $1^+$ state that receives most of the total width, for heavier nuclei $\Gamma_\gamma$ is fragmented into several states. We have checked that, as expected, electric quadrupole transitions are roughly two orders of magnitude smaller than the $M1$ ones presented in Table~\ref{tab:M1_single-gamma_widths}.

\begin{table}[t]
\begin{center}
\def\arraystretch{1.3}
\begin{tabular}{c c c c c } 
\hline\hline
  Nucleus  & $ H_{\rm{eff}}$  & $ E_{1^+}(\mathrm{MeV})$ & $\Gamma_\gamma (\text{eV})$ & BR(\%)  \\ 
\hline
  $^{48}$Ti  &  KB3G  &  12.2  &  6.8   &  92 \\
\hline  
  $^{48}$Ti  &  GXPF1B  &  12.6  &  13  &  97 \\
\hline  
\multirow{4}{*}{$^{76}$Se}  &  \multirow{4}{*}{GCN2850}  & 12.0 &  0.77  &  24\\
& &  12.3  &  0.47  &  15 \\
& &  12.8  &  0.63  &  20 \\
& &  13.2  &  1.1   &  35 \\
\hline  
\multirow{2}{*}{$^{76}$Se}  &  \multirow{2}{*}{JUN45}  &  12.9  &  2.7   &  68 \\
& &  14.1  &  0.96   & 25 \\
\hline  
\multirow{3}{*}{$^{76}$Se}  &  \multirow{3}{*}{JJ44B} &  13.4  &  0.60  &  14 \\
& &  13.5  & 1.7   & 41 \\
& &  15.4   & 0.96   & 23 \\
\hline  
\multirow{2}{*}{$^{82}$Kr}  &  \multirow{2}{*}{GCN2850} & 13.1 & 2.3 & 36 \\
& & 14.9 & 2.9 & 45 \\
\hline  
\multirow{2}{*}{$^{82}$Kr}  &  \multirow{2}{*}{JUN45}  & 13.7 & 1.7 & 34 \\
& & 14.6 & 1.3  & 27 \\
\hline  
\multirow{3}{*}{$^{82}$Kr}  &  \multirow{3}{*}{JJ44B} & 19.3 &  0.61 &  25 \\
& &  19.6  &  0.66  &  27 \\
& &  21.9  &  0.82  &  34 \\
\hline  
\multirow{3}{*}{$^{130}$Xe}  &  \multirow{3}{*}{GCN5082}   & 15.9 &  0.42 &  15 \\
&  &  17.9 &  1.1  &  39 \\
&  &  18.4  &  0.45  &  16 \\
\hline  
\multirow{4}{*}{$^{136}$Ba}  &  \multirow{4}{*}{GCN5082}  & 17.5 &  1.2 &  10 \\
& &  17.9  &  2.2  &  18 \\
& &  19.0  &  4.4  &  36 \\
& &  19.3  &  1.5  &  12 \\
    \hline\hline
\end{tabular}
\end{center}
\caption[]{Partial decay widths ($\Gamma_{\gamma}$), branching ratios (BR) and energies ($E_\gamma$) of the dominant $M1$ $0^+_\text{DIAS}\rightarrow 1^+_{n}$ transitions calculated with the nuclear shell model using different interactions ($H_{\rm{eff}}$).}
\label{tab:M1_single-gamma_widths}
\end{table}

\subsection{Proton emission}
DIAS generally lie above single-proton, $S_p$, and single-neutron, $S_n$, separation energies, opening nucleon-emission branches. We focus on the proton emission from the DIAS:
\begin{equation} ^{A}_{Z}Y_N^\text{DIAS}\longrightarrow\, ^{A-1}_{Z-1}X_N^*+p\,,
\end{equation}
permitted if $Q_p=B(N,Z)+E_{\rm{DIAS}}-B(N,Z-1)-E^\text{exc}_f>0$, with $B$ the nuclear binding energy and $E^{\rm{\small{exc}}}_f$ the excitation energy of the final state. For the case of $^{48}$Ti, this condition forbids the proton emission to the $7/2^-$ IAS in $^{47}$Sc ($Q_p=-2.47\rm MeV$), and we expect this situation to also hold in heavier nuclei where $E_\text{DIAS}$ is not known. In contrast, the low-lying states of the final nucleus, with $T_f=T_{GS}+1/2$, are energetically-favored. Thus, from the DIAS isospin changes in $\Delta T=3/2$, and these transitions can only proceed via isospin-symmetry breaking. In fact, similar transitions from IAS are a valuable tool to study isospin mixing in nuclei~\cite{Brown_1990,Smirnova_2017}.

The decay width of the proton-emitting state can be factorized as~\cite{Macfarlane60,Aberg} 
\begin{align}
\Gamma_p=\Gamma_{sp}\,S\,,    
\end{align}
where $S$ is the spectroscopic factor which represents the proton preformation amplitude, and $\Gamma_{sp}$ is known as the single-particle width. We compute  the latter using the Wentzel-Kramers-Brillouin (WKB) method~\cite{Macfarlane60,Aberg,Buck_1992,Poenaru_1989}, but for the proton emission to states where this semiclassical approach is not applicable we use the code \texttt{wspot}~\cite{ABrown}, which calculates the widths of unbound resonances. Both approaches agree very well whenever the WKB method can be applied.

\begin{table}[t]
\begin{center}
\def\arraystretch{1.3}
\begin{tabular}{c c c c c}
\hline\hline
$Q_p(\mathrm{MeV})$ & $J^{\pi}$  & $\Gamma^{\rm{emp}}_{p}(\rm{eV})$ & $\Gamma^{\rm{Coul}}_{p}(\rm{eV})$ & $\Gamma^{\chi}_{p}(\rm{eV})$  \\
\hline
 5.93 & $\frac{7}{2}^-$ & 19 & 2.9 & 33 \\
\hline
5.13  &$\frac{3}{2}^-$  & 59 & 46 & 67 \\
\hline
4.64 & $\frac{5}{2}^-$  & 2.3 & 7.0 & 14 \\
\hline
3.74 & $\frac{1}{2}^-$  & 19 & 70 & 95 \\
\hline
$>0$ & all & 130  & 130 & 280 \\
\hline\hline
\end{tabular}
\end{center}
\caption{Proton emission widths from the DIAS in $^{48}$Ti to the lowest-lying states in $^{47}$Sc with a given angular momentum and parity, $J^{\pi}$. The results combine single-particle widths calculated with the WKB method and \texttt{wspot}~\cite{ABrown} with spectroscopic factors obtained semi-empirically ($\Gamma_p^\text{emp}$) and with isospin-breaking Coulomb ($\Gamma_p^\text{Coul}$) and chiral ($\Gamma_p^\chi$) interactions (see the text for details). The total proton emission width to all energetically allowed states is also given.}
\label{tab:proton_widths}
\end{table}

Since our shell-model interactions are isospin symmetric, the corresponding spectroscopic factor for the $\Delta T = 3/2$ proton emission vanishes. To overcome this limitation, we estimate $S$ in three different ways. Firstly, we use the measured proton emission width from the  DIAS in $^{32}$S to the $^{31}$P GS, which is also a $\Delta T = 3/2$ emission, $\Gamma_p=38\rm~eV$~\cite{Wilkerson,Endt}. Since theoretically we obtain $\Gamma_{sp}=1.05\rm~MeV$, we extract a semi-empirical spectroscopic factor, $S^{\rm{emp}}=3.7\times 10^{-5}$. Alternatively, we calculate $S$ using shell-model interactions that break isospin symmetry. We first focus on the proton emission of $^{48}$Ti. Here we use KB3G supplemented with the Coulomb interaction, obtaining $S^{\rm Coul}=6.5\times 10^{-6}$  for the decay to the $^{47}$Sc GS. In addition, we also obtain the spectroscopic factor calculating both $^{48}$Ti and $^{47}$Sc within the valence-space similarity renormalization group (VS-IMSRG) method, with interactions based on a chiral effective field theory Hamiltonian with three-nucleon forces, labeled EM1.8/2.0~\cite{Entem_2003,Bogner_2010,Hebeler2011}. The interactions are obtained with an initial VS-IMSRG decoupling in a harmonic-oscillator space truncated to $e=2n+l\leq12$, and besides three-nucleon forces are truncated to $e_1+e_2+e_3\leq24$. In turn, operators are truncated to the normal-ordered two-body level, VS-IMSRG(2). This approach gives $S^{\chi}=6.3\times 10^{-5}$ for the proton emission to the GS. Therefore, for the purpose of this work, the three determinations of the spectroscopic factor are rather consistent. 

The results in Table~\ref{tab:proton_widths} combine these spectroscopic factors with the calculated single-particle widths to predict the proton emission from the DIAS in $^{48}$Ti to the lowest-energy states in $^{47}$Sc. In addition, Table~\ref{tab:proton_widths} also shows the results for the total proton emission width to all energetically allowed states in $^{47}$Sc.
We obtain $\Gamma_p^\text{emp}$ using the same semi-empirical $S^{\rm{emp}}$ across all possible proton emissions from the $^{48}$Ti DIAS. However, for the widths associated with the isospin-breaking KB3G-plus-Coulomb and chiral interactions we compute $S^\text{Coul}$ and $S^\chi$ for the emission to each final state.

The main conclusion of Table~\ref{tab:proton_widths} is that the $\Delta T =3/2$ proton-emission width is about an order of magnitude larger than $\Gamma_\gamma$, so that we expect the former to dominate the decay of the DIAS.
In general, the emission to the states with higher $Q_p$ dominates each $J^\pi$ combination, being above $80\%$ of all contributions. However, in the VS-IMSRG calculation for $3/2^-$ there is a $\sim50\%$ contribution to higher-energy states because they present a larger $S^\chi$.

Table~\ref{tab:all_channels_BRs} summarizes the partial branching ratios for all decay processes considered. In order to give indicative estimated values, we average the $\Gamma_{\gamma\gamma}$ and $\Gamma_{\gamma\gamma}/\Gamma_{\gamma}$ results over the shell-model interactions used in Tables~\ref{tab:M1M1_gamma-gamma_widths} and~\ref{tab:M1_single-gamma_widths}. For the $\Gamma_p$ of $^{48}$Ti we also average over the three different values obtained, reflecting the different spectroscopic factors. In contrast, for heavier nuclei, we assume the semi-empirical spectroscopic factor and consider a limited number of low-lying states in the final nucleus, thus representing an upper bound to $\Gamma_{\gamma\gamma}/\Gamma_{p}$. As discussed earlier, for all nuclei the competing processes, especially proton emission, are much more likely to occur than the $\gamma\gamma$-$M1M1$ decay. The $\Gamma_{\gamma\gamma}/\Gamma_{\gamma}$ branching ratio in $^{48}$Ti is very similar to the one found for the lowest $2^+$ state in $^{48}$Ca in Ref.~\cite{Severyukhin:2021kou}.  

\begin{table}[t]
\begin{center}
\def\arraystretch{1.3}
\begin{tabular}{c c c c l} 
\hline\hline
   Nucleus  & $\Gamma_{\gamma\gamma}/\Gamma_{\gamma}$ & $\Gamma_{\gamma\gamma}/\Gamma_{p}$  & $\Gamma_{\gamma\gamma}/\Gamma_{\mathrm{IPC}}$ & $\Gamma_{\gamma\gamma}/\Gamma_{\mathrm{IC}}$\\ 
\hline
  $^{48}$Ti   & $2\times 10^{-8}$  & $\quad7\times 10^{-10}$  &  $3\times 10^{-6}$  &  0.06  \\
\hline
  $^{76}$Se   & $5\times 10^{-7}$  & $<4\times 10^{-9}$  &  $7\times 10^{-6}$  &  0.07  \\
\hline
  $^{82}$Kr   & $7\times 10^{-7}$  & $<7\times 10^{-9}$  &  $3\times 10^{-6}$  &  0.03  \\
\hline
  $^{130}$Xe  & $8\times 10^{-8}$  & $<6\times 10^{-9}$ &  $1\times 10^{-7}$  &  $0.0002$  \\
\hline
  $^{136}$Ba  & $1\times 10^{-7}$ & $<3\times 10^{-9}$  &  $6\times 10^{-7}$  &  0.001  \\  
\hline\hline
\end{tabular}
\end{center}
\caption[]{Predicted branching ratios of the $\gamma\gamma$-$M1M1$ decay of the $0^+_\text{DIAS}$ with respect to the competing decay channels: single-$\gamma$ $M1$ decay, proton ($p$) emission, internal pair creation (IPC) and internal conversion (IC), for all studied nuclei. See the text for details.}
\label{tab:all_channels_BRs}
\end{table}

\section{Strategy for the measurement of double-magnetic dipole transitions from DIAS}
To conduct precise $\gamma$-ray measurements, high-purity germanium (HPGe) arrays represent a natural choice, given their unparalleled energy resolution.
However, neither previous attempts using traditional HPGe arrays such as GAMMASPHERE~\cite{dgdGAMMASPHERE} nor feasibility studies~\cite{brugnara} with state-of-the-art HPGe arrays such as AGATA~\cite{AGATA,AGATA_LNL} have succeeded in measuring $\gamma\gamma$ transitions.
An alternative solution involves using LaBr$_3$ detectors, which present good energy resolution and excellent time resolution.
These characteristics proved to be crucial to discriminate $\gamma\gamma$ from background events in the experiments performed by Walz $et$ $al.$ \cite{Walz15} and by Söderström $et$ $al.$~\cite{Söderström2020}. 

Considering this, we have developed a GEANT4~\cite{AGOSTINELLI2003250} simulation to characterize the contributions of the competing processes and determine the optimal data processing methods so as to maximise the $\gamma\gamma$ detection efficiency. In particular, we simulate the experimental setup and its response to the radiation emitted in the nuclear de-excitation of the DIAS in $^{48}$Ti~\cite{Thesis_Damiano}. The  simulated setup consists of 302 3.5''x8'' LaBr$_3$ cylindrical scintillator detectors in spherical configuration, placed 50 cm away from the center of the sphere. A 1 mm thick aluminium capsule for the crystals was included, on top of a 2 mm thick scattering chamber, concentric with the array of detectors and also in aluminium. This geometry is dictated by the necessity of having a large angular acceptance array capable of detecting very energetic $\gamma$ rays with high efficiency. 

\begin{figure}[t]
    \centering
    \includegraphics[width=
    0.45\textwidth]{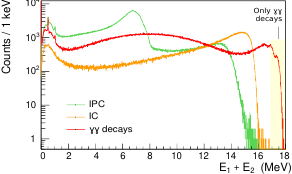}
\caption{Simulated energy spectrum of the energy sum between the two hits acquired in the setup, for $\gamma\gamma$ decay (red curve), internal pair creation (IPC, green) and internal conversion (IC, orange) events. }
    \label{fig:processes1}
\end{figure}

\subsection{Discrimination from competing processes}
Since the decay widths of the competing channels are orders of magnitude larger than $\Gamma_{\gamma\gamma}$, it is essential to find a set of observables that allows one to discriminate between the competing processes and the $\gamma\gamma$ decays of interest. 
In particular, we focus on processes leaving $^{48}$Ti in the GS, so that the energy of the emitted radiation matches that of the $\gamma\gamma$ decay. Indeed all the remaining competing processes, including proton decay, can be easily discriminated from the $\gamma\gamma$ decay events by looking at the total energy deposited in the detectors.

In contrast, the discrimination between $\gamma\gamma$ and other electromagnetic processes requires a careful study. We simulate separately 10$^7$ events for each of these processes and analyze the detectors output. In this preliminary stage of the data analysis, we select exclusively multiplicity 2 events, i.e. when only two detectors are hit, considering $^{48}$Ti to be excited at rest. 

\subsubsection{Internal conversion and internal  pair creation}
For the charged leptons generated in these processes, we expect an energy loss of about 2 MeV in the scattering chamber and detector capsules material. For this reason, the energy sum of the two hits detected, $E_1+E_2$, can be employed as the discriminating factor between the IPC, IC and $\gamma\gamma$ decays. Figure~\ref{fig:processes1} shows the corresponding spectrum for these three channels, and illustrates that at high energies close to $E_\text{DIAS}$ only $\gamma\gamma$ decays contribute. 


\subsubsection{First-order electromagnetic transitions}  
\begin{figure}[t]
    \centering
    \includegraphics[width=
    0.46\textwidth]{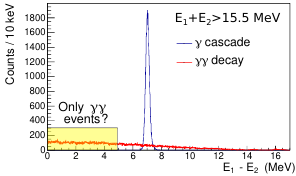}
\caption{
Simulated spectrum of the absolute energy difference between the two hits acquired in the setup, considering only cases in which their energy sum is larger than 15.5~MeV, for $\gamma\gamma$ decay (in red) and $\gamma$ cascade (in blue) primary events. }
    \label{fig:processes2}
\end{figure}

The $^{48}$Ti DIAS de-excitation through an intermediate state, resulting in a $\gamma$-$\gamma$ cascade to the ground state, has a very large decay width $\Gamma_{\gamma}$ compared to the $\gamma\gamma$ decay one: $\Gamma_{\gamma\gamma}/\Gamma_{\gamma}\approx 10^{-8}$, see Table~\ref{tab:all_channels_BRs}. The main difference between the two channels is that in the $\gamma\gamma$ decay the two $\gamma$ rays are emitted with a continuous energy spectrum, which, in contrast, is discrete for the $\gamma$-$\gamma$ cascade. It is therefore natural to employ the absolute energy difference between the two hits, $E_1-E_2$, as the discriminating factor between the two processes. Figure~\ref{fig:processes2} shows the corresponding spectrum for both decay channels, considering only events with energy sum $E_1+E_2>15.5$~MeV, and a de-excitation through the 1$^+$ state with 12.35~6 MeV excitation energy for the $\gamma$ cascade---this is the most likely transition according to our calculations, see Table~\ref{tab:M1_single-gamma_widths}. Since only $\gamma\gamma$ events carrying roughly the same energy are relevant for the correlation with $0\nu\beta\beta$ decay, one could, in principle, focus in the region $|E_1-E_2|\approx0$, highlighted in Fig.~\ref{fig:processes2}. 


\begin{figure}[t]
    \centering
    \includegraphics[width=
    0.45\textwidth]{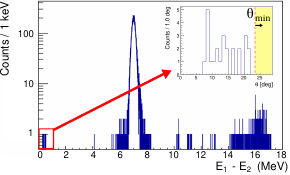}
\caption{Same as Fig.~\ref{fig:processes2} for $\gamma$ cascade primary events, but with higher resolution in the y axis. The inset presents the low-energy spectrum for the angular correlation between the two hits. }
    \label{fig:processes3}
\end{figure}

However, imposing these energy conditions is not sufficient to isolate $\gamma\gamma$ decays getting rid of all $\gamma$-cascade events. Indeed, Figure~\ref{fig:processes3} highlights that, still, few $\gamma$-cascade counts lie in the $|E_1-E_2|<1$~MeV region. These correspond to events where the 12.356~MeV primary photons deposit only about 8.7~MeV in the first crystal hit, and the remaining $\approx4$~MeV, carried by highly energetic secondary radiation, leave this region and are absorbed in the crystal in which the other primary photon is detected. Our simulations indicate a possible solution to this problem exploiting both the positional and temporal information of the hits: the secondary radiation takes some time to reach the second crystal after the primary photons hit. Therefore, for large enough correlation angles between the two hits, before reaching the second crystal the radiation travels for a longer time than the LaBr$_3$-detectors time resolution, and its energy will not be deposited with the second hit. In this way, considering only  events with a correlation angle  $\theta>\theta_\text{min}$, besides the energy conditions, $E_1+E_2>15.5$~MeV and $|E_1-E_2|<1$~MeV, it is possible to get rid of all the simulated $\gamma$-decay cascade events. The inset of Figure~\ref{fig:processes3} illustrates this strategy, which follows essentially the same idea used by Walz $et$ $al.$ to distinguish between prompt $\gamma\gamma$ events and Compton-scattered primary photons in the competitive $^{137}$Ba decay experiment \cite{Walz15}. The excellent time resolution of LaBr$_3$ turned out to be crucial for this discrimination, justifying their use also in this study.


\section{Summary and conclusions}

Nuclear matrix elements are crucial to analyze $0\nu\beta\beta$-decay experiments, but they are currently poorly known. Recent studies have proposed that measuring 
$\gamma\gamma$-$M1M1$ transitions from DIAS can provide insights on the values of these $0\nu\beta\beta$-decay matrix elements. In this article, we provide detailed calculations of the $\gamma\gamma$-$M1M1$ widths,  $\Gamma_{\gamma\gamma}$, obtained with the nuclear shell model for several nuclei used in $0\nu\beta\beta$-decay searches.
We also quantify the decay widths of the main competing processes: proton emission, single-$\gamma$ decay, internal conversion and internal pair creation. We complement these theoretical calculations by proposing a strategy to distinguish experimentally the desired $\gamma\gamma$-$M1M1$ transitions from these competing processes, based on using detector setups like LaBr$_3$ scintillator arrays. Even though the decay width of the $M1M1$ decay is expected to be suppressed by several orders of magnitude with respect to the proton-emission and $\gamma$-decay branches, our simulations indicate that these $\gamma\gamma$ transitions may be detected. This work opens an avenue to inform about $0\nu\beta\beta$ nuclear matrix elements using $\gamma$ spectroscopy.


\section*{Acknowledgements}
We would like to thank Prof. T. Kibedi for the help with the IC and IPC calculations.
This work was supported by the
MCIN/AEI/10.13039/5011
00011033 from the following grants: PID2020-118758GB-I00, PID2023-147112NB-C22, RYC-2017-22781 through the “Ram\'on
y Cajal” program funded by FSE “El FSE invierte en tu futuro”, CNS2022-135716 funded by the
“European Union NextGenerationEU/PRTR”, and CEX2019-000918-M to the “Unit of Excellence Mar\'ia de Maeztu 2020-2023” award to the Institute of Cosmos Sciences; by the Generalitat de Catalunya, grant 2021SGR01095 and by the U.S. Department of Energy under Contract No. DE-FG02-97ER4101.

\appendix



\bibliographystyle{apsrev4-1} 
\bibliography{bibliography.bib}






\end{document}